\documentclass[conference]{IEEEtran}

\usepackage[letterpaper, left=1in, right=1in, bottom=1in, top=0.75in]{geometry}
\usepackage[utf8]{inputenc}\usepackage{amsmath}
\usepackage{graphicx}
\usepackage{tikz-cd}
\usepackage{tikz}
\usetikzlibrary{calc}

\usepackage{listings}
\usepackage{color}
\definecolor{dkgreen}{rgb}{0,0.6,0}
\definecolor{gray}{rgb}{0.5,0.5,0.5}
\definecolor{mauve}{rgb}{0.58,0,0.82}
\title{Trustless Price Feeds of Cryptocurrencies: Pathfinder}
\author{\IEEEauthorblockN{Orhan Koc}
    \IEEEauthorblockA{hankoc@onchainfactory.com} 
    \IEEEauthorblockA{May 21, 2023} 
    }

\begin{document}
    \maketitle
    \begin{abstract}
        Price feeds of securities is a critical component for many financial services, allowing for collateral liquidation, margin trading, derivative pricing and more. With the advent of blockchain technology, value in reporting accurate prices without a third party has become apparent. There have been many attempts at trying to calculate prices without a third party, in which each
        of these attempts have resulted in being exploited by an exploiter artificially inflating the price. The industry has then shifted to a more centralized design, fetching
        price data from multiple centralized sources and then applying statistical methods to
        reach a consensus price. Even though this strategy is secure compared to reading from a single source, enough number of sources need to report to be able to apply statistical
        methods. As more sources participate in reporting the price, the feed gets more secure with the slowest feed becoming the bottleneck for query response time, introducing a tradeoff between security and speed. This paper provides the design and implementation details of a novel method to
        algorithmically compute security prices in a way that artificially inflating targeted pools has
        no effect on the reported price of the queried asset. We hypothesize that the proposed
        algorithm can report accurate prices given a set of possibly dishonest sources.
    \end{abstract}

    \section{Introduction}
        There have been many attempts at calculating cryptocurrency prices without the need of
        a third party by fetching price from Automated Market Makers (AMMs). Depending on the
        oracle design, the strategy to fetch or calculate the price can become a bottleneck
        for decentralized and trustless systems. AMM-based oracles in the past have not been
        a reliable way of providing price data mainly due to the permissionless nature of AMM
        pools, where exploiters can easily manipulate the price of targeted pools. There have been numerous attempts at calculating price feeds without the need of a
        third party. Some of the more prominent techniques of such pricing algorithms and
        their drawbacks are explained in this section.

        \textbf{Spot Pool Price} Decentralized Exchanges (DEX) like Uniswap are Automated Market Makers (AMMs) in the sense that the trader executes trades against a pool with a mathematical relationship between the two assets, bound by liquidity. For every asset pair there exists a pool with reserves for corresponding assets. One approach at algorithmically reporting the price of an asset is to calculate the ratio of the reserves; this approach yields the price of the pair for that specific pool.
        
        Using Spot Pool Price is a relatively primitive method of calculating the price of
        an asset and assumes the pool reflects the consensus on value accurately. This
        approach is prone to exploits given the pools are permissionless in nature, so the price of a pair can be changed momentarily by anyone.
        
        \textbf{Schelling Game}
        Schelling Game is basically stake-based voting to determine the price of an asset,
        and assumes cryptoeconomic security by slashing voters that posted outliers.
        Schelling method is subject to whale manipulation with a varying corruption cost based
        on token distribution. 

        \textbf{Time Weighted Average Price} Given blockchains are essentially ledgers, historical price can be kept available for processing. TWAP leverages this by assigning weights to prices with respect to occurence, resulting in the average price of a security over a specified time. The attacker needs to “hold“ the price until the next checkpoint, making the cost of exploit unpredictable and incurring heavy losses in the presence of arbitrageurs trying to balance the market.

        Using a TWAP, as demonstrated by Uniswap and Keep3r, helps mitigate the noise by
        short-lived price hikes by implementing a simple moving average. This method introduces a tradeoff between data freshness and security, as it scales inversely with recency.

    \section{Exploit Review}
        \textbf{Oracle:} Kyber

        \textbf{Target Application:} bZx Finance  

        \textbf{Industry:} Lending  

        \textbf{Loss:} \textdollar 670,000  

        \textbf{Summary:} Borrow via flashloan, Pump sUSD by selling ETH/sUSD on Kyber, Borrow
        using sUSD as collateral (with inflated sUSD price), Pay back flashloan and the
        position is underwater.

        \medskip

        \textbf{Oracle:} Curve

        \textbf{Target Name:} Harvest Finance

        \textbf{Industry:} Yield Aggregator

        \textbf{Loss:} \textdollar 24M

        \textbf{Summary:} The attacker takes out a flashloan of USDT and USDC, convert USDT to
        USDC on Curve pool to inflate the price, deposit USDT to Harvest finance in exchange
        for pool shares, sell USDT vs USDC to push down prices, withdraw pool shares.

        \medskip

        \textbf{Oracle}: Curve

        \textbf{Target Name:} Value DeFi

        \textbf{Industry} Yield Aggregator

        \textbf{Loss} \textdollar 7.4M

        \textbf{Summary}: Value DeFi uses Curve pools to determine the price of assets. The
        attacker manipulated the marginal price of an asset in the Curve pool using a
        flashloan from Aave.

        \medskip

        \textbf{Oracle}: Uniswap

        \textbf{Target Name}: Warp Finance

        \textbf{Industry}: Lending

        \textbf{Loss} \textdollar 7.8M

        \textbf{Summary}: The attacker manipulates the price of the Uniswap pool with a flash
        loan to manipulate the price of DAI and borrows against DAI. Warp Finance fetches
        price via single Uniswap pool.

        \medskip

        \textbf{Oracle}: Uniswap

        \textbf{Target Name}: Cheese Bank

        \textbf{Industry}: Fund management

        \textbf{Loss} \textdollar 3.3M

        \textbf{Summary}: Attacker inflates the price of his collateral via flashloan pump of
        Uniswap's pool.

        \medskip

        \textbf{Oracle}: Keep3r

        \textbf{Target Name}: Inverse Finance

        \textbf{Industry}: Lending

        \textbf{Loss} \textdollar 15M

        \textbf{Summary}: Attacker manipulates INV/ETH price on SushiSwap, borrows using INV
        as collateral. Inverse Finance uses Keep3r as an oracle service, which uses a Time
        Weighted Average Price (TWAP), and the attacker was able to inflate his collateral.

        \medskip

        \textbf{Oracle}: Muon Network

        \textbf{Target Name}: Deus Finance

        \textbf{Industry}: Synthetic Asset Platform

        \textbf{Loss} \textdollar 3M

        \textbf{Summary}: flash-loan-assisted manipulation of price oracle that reads the
        price from the pair of StableV1 AMM - USDC/DEI

        \medskip

        
        We will use these examples to devise the attack-vector in its abstraction to better
        understand the problem of trustless price feeds. Using AMM-based-oracles as an attack
        vector is generally carried out in the following steps:

        \begin{enumerate}
            \item Find the pool observed by the oracle to fetch \textdollar ABC price.
            \item Buy \textdollar ABC from the target pool in a large sum to inflate the
            price.
            \item Use \textdollar ABC as collateral to get a loan. OR deposit \textdollar ABC
            to a vault to get proprietary tokens
            \item \textdollar ABC price stabilizes, and the position becomes underwater.
        \end{enumerate}

        The strategy to acquire liquidity to execute these attacks can be categorized into
        capital-intensive and flashloan attacks. A capital-intensive attack requires an
        attacker to risk their funds to manipulate the price of an asset. A flashloan attack
        requires attackers to “prove” {loanAmount + interest + gas} will be smaller than the profit, and can be paid in the same transaction. The attacker only needs to fund the gas for the transaction, and the
        protocol will provide the capital as a flashloan. This type of sourcing significantly reduces the
        monetary barrier to executing an attack.  

    \section{Algorithm Design}
        Ultimately, the goal is to calculate the price of any cryptocurrency without trusting
        a third party, accounting for manipulated price sources. This challenge can be broken down into two stages:
        \begin{itemize}
            \item Quantify the degree of manipulation for a given pool.
            \item Calculate the price of a pair using honest sources.
        \end{itemize}
        
        \subsection{Basic Setting}
            
            \textbf{Calculating Price:} \textit{Absolute Price} is defined as the equilibrium point of the demand and supply for an asset; where the demand is denoted in terms of USD. When the demand is denoted in terms of another good, which is the case for most foreign exchange and cryptocurrency markets, the price is referred
            to as the \textit{Relative Price}. 

            It's important to note \textit{Absolute Price} of any asset can be represented in terms of compatible \textit{Relative Asset} pairs. The simplest example for this calculation is
            multiplying the price of BTC/ETH with ETH/USD to find the price of BTC/USD. In abstraction, we can
            calculate the price of any asset, $P_{A/E}$ or price of asset $A$ in asset $E$, using n asset pairs such that asset $A$ is the numerator of pair $0$,
            asset $E$ is the denominator of pair $n$, with a set of pairs $1:n-1$ which the
            denominator of the prior is the numerator of the following. This queue of pairs is compatible in the sense that the multiplication of these pairs, if the markets are efficient, should equal BTC/USD. The visualization of a compatible queue in its abstraction can be seen in Figure 1.

            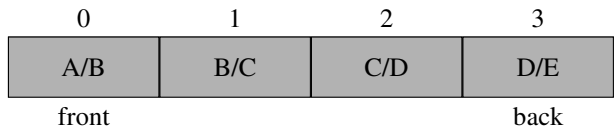
\begin{figure}[h!]
                \begin{center}
                    \def\clap#1{\hbox to 0pt{\hss#1\hss}} \def\cells#1#2#3{%
                    \coordinate (next) at (0,0); \foreach [count=\i from 0] \j  in {1,...,#1} {
                    \node[cell,label=above:\i,anchor=west] (cell\i) at (next) {}; \coordinate (next)
                    at (cell\i.east); }

                    \pgfmathsetmacro{\last}{#3+#2-1}
                    \foreach \i in {#3,...,\last} {
                        \pgfmathsetmacro{\back}{int(mod(\i,#1))}
                        \node[shaded cell] (back) at (cell\back) {}; }

                    \node[below] at (cell#3.south) {\clap{front}}; \node[below] at (back.south)
                    {\clap{back}}; \node at (cell0) {A/B}; \node at (cell1) {B/C}; \node at (cell2)
                    {C/D}; \node at (cell3) {D/E}; }

                    \tikzset{ cell/.style = {draw, minimum width=20mm, minimum height=0.8cm}, shaded
                    cell/.style = {cell, fill=black!30}, }

                    \begin{tikzpicture}
                    \cells{4}{4}{0}
                    \end{tikzpicture}
                    
                \end{center}
                \caption{Queue of Compatible Pairs}
                \label{fig:birds}
            \end{figure}
            
            \textbf{Quantifying manipulation:} A manipulated pool by definition is a pool that doesn't reflect the price of an asset accurutely. In other words, it is a pool with a relatively high arbitrage difference  compared to the rest of the market. Out of the potential arbitrage opportunities this imbalance creates, the triangular arbitrage potential is a well established tool in stating the irregularity with respect to other asset pairs. Using this logic of relative quantification of pool manipulation, we can describe a collection of Decentralized Exchange pools using an undirected, cyclic graph where the nodes are added in the granularity of  triangular arbitrage triangles. The general form of an arbitrage triangle is comprised of pairs $p_1$: A / B, $p_2$: B / C, $p_3$: C / A. An example arbitrage triangle is EUR/YEN, YEN/USD, USD/EUR. 

            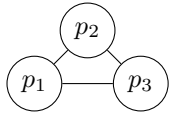
\begin{figure}[h!]
                \begin{center}
                    \begin{tikzpicture}[main/.style = {draw, circle}] 
                        \node[main] (1) {$p_1$}; \node[main] (2) [above right of=1] {$p_2$};
                        \node[main] (3) [below right of=2] {$p_3$}; \draw[-] (1) -- (2); \draw[-]
                        (2) -- (3); \draw[-] (3) -- (1);
                    \end{tikzpicture} 
                \end{center}
                \caption{Arbitrage Triangle}
                \label{fig:trio}
            \end{figure}

            Figure 2 shows the said modules, comprised of Decentralized Exchange pools as nodes $p_1$, $p_2$, $p_3$ that equal $1$ when multiplied in an efficient market. It's important to note this form of pool queue is only slightly different than the compatible queue proposed under the definition of Relative Price. The difference of an arbitrage queue is that numerator of source pair should be the same as the denominator of target pair such that the multiplication of pool prices should equal 1. It's important to note the arbitrage cycle can be longer than three nodes as shown in Figure 2. Circular nature of arbitrage queue with 7 pairs, where successive nodes share an asset can be seen in Figure 3. 
            
            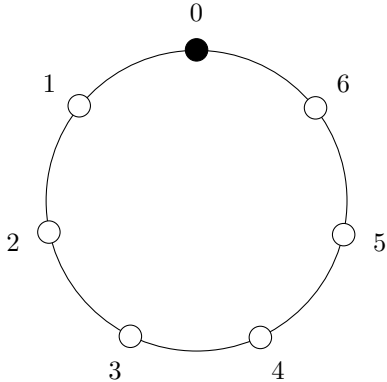
\begin{figure}[h!]
                \begin{center}
                    \begin{tikzpicture}
                    \def\numofpoints{7}
                    \def\circpatt{{1,0,0,0,0,0,0}} 
                    \def\labelpatt{{0,1,2,3,4,5,6}}
                    
                    \node[circle,draw,minimum width=4cm] (bigc) {};
                    \foreach \x in {0,...,\numexpr\numofpoints-1\relax}
                    {
                    \pgfmathparse{\circpatt[\x]}
                    \ifnum\pgfmathresult>0\relax\def\mycolor{black}\else\def\mycolor{white}\fi
                    \node[circle,inner sep=3pt,draw,fill=\mycolor] (n-\x) at (bigc.360/\numofpoints*\x+90) {};
                    \node (l-\x) at (360/\numofpoints*\x+90:2.5cm)  {\pgfmathparse{\labelpatt[\x]}$\pgfmathresult$};
                    }
                    \end{tikzpicture}
                \end{center}
            \caption{Arbitrage Queue}
            \label{fig:loop}
            \end{figure}
            
            Regardless of how long the arbitrage queue is, the product $P$ of $n$ pools that form an arbitrage triangle at time $t$ can be denoted as: 

            \[\prod_{n=0}^{} p_{n,t} =  P_{t}\] 

            In regular market conditions, markets aren't perfectly correlated due to differences in demand and liquidity. So in reality $P_{t} \neq 1$ and arbitrageurs will start to take profit when the price difference is high enough to at least cover the cost of executing the transaction bundle. Given pools are permissionless, and an arbitrage queue that deviates enough from $1$ will be reduced closer to $1$ for certain,
            we can certainly say that the price of assets will converge to their fair shares.

            \[\lim_{t\to\infty } P_{t}  = 1 \]
            
            Another way to look at the product of this price sequence is that the deviation from 1  in any direction, 
            \[ I_t = |1-P_t| \]
            would quantify an irregularity in a pair trio and is expected to converge to 0 in each block of transactions because arbitrageurs compete with each other.

            $I_t$ only conveys information as to whether irregularity occurs in
            the triangle consisting of three pairs; On its own, this form of
            irregularity quantification doesn't convey which of the three pairs
            is causing the imbalance. If, however,  arbitrage triangles with
            two other pairs are added for each node, such that each pair in the
            triangle is connected to one other unique arbitrage triangle as
            shown in Figure 4, then the overarching graph quantifies relative
            irregularity in pair granularity. The arbitrage triangles in Figure
            4 are $p_7$:$p_4$:$p_6$ as triangle 1, $p_1$:$p_3$:$p_5$ as
            triangle 2 and $p_2$:$p_9$:$p_8$ as triangle 3.
            
            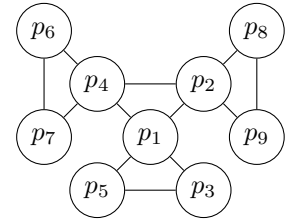
\begin{figure}[!h]
                \begin{center}             
                    \begin{tikzpicture}[main/.style = {draw, circle}] 
                    \node[main] (1) {$p_1$}; 
                    \node[main] (2) [above right of=1] {$p_2$};
                    \node[main] (3) [below right of=1] {$p_3$}; 

                    \node[main] (4) [above left of=1] {$p_4$};
                    \node[main] (5) [below left of=1] {$p_5$}; 

                    \node[main] (6) [above left of=4] {$p_6$};
                    \node[main] (7) [below left of=4] {$p_7$}; 

                    \node[main] (8) [above right of=2] {$p_8$};
                    \node[main] (9) [below right of=2] {$p_9$}; 

                    \draw[-] (1) -- (5);
                    \draw[-] (1) -- (3);
                    \draw[-] (5) -- (3);

                    \draw[-] (6) -- (4);
                    \draw[-] (4) -- (7);
                    \draw[-] (6) -- (7);

                    \draw[-] (2) -- (8);
                    \draw[-] (9) -- (8);
                    \draw[-] (9) -- (2);

                    \draw[-] (1) -- (2);
                    \draw[-] (4) -- (2);
                    \draw[-] (1) -- (4);

                    \end{tikzpicture} 
                \end{center}
                \caption{Intorconnected Arbitrage Triangles}
            \end{figure}

            Now that we have established the building blocks behind constructing nodes and vertices, for $n$ different arbitrage triangles we denote Graph $G$ at time $t$ as:

            \[ G_t = G_{1,t} \cup G_{2,t} \cup G_{3,t} ... \cup G_n \] 
            
            with arbitrage triangles, $G_n$, consisting of a set of compatible pairs, $V$, have corresponding edge weights $I_{n,t}$:

            \[ G_{n,t} = (V, I_{n,t}) \]

        \subsection{Constructing the Graph}
            The building block of this graph are the arbitrage triangles with their corresponding edge weights. In order to create this framework using available on chain data, we will be recording the price of each pool as \(p = asset1/asset2\) on each decentralized exchange and its corresponding pool price. For each pair in the list of asset pairs, a unique arbitrage triangle is searched. If a pair trio can be constructed as shown in Figure 2, the deviation $I_t$ for the trio is calculated and entered as the edge weights between the pairs.

            This procedure is repeated for each pair in the list of pairs, skipping triangles that are already added to the graph. It's important to note the order of numerator and denominator is not important as the reciprocals of the price can be taken to adjust the price to the original form. Meaning if we have BTC/USDT and USDT/ETH pairs available in the list, then both ETH/BTC and BTC/ETH would work as a third pool if the pool price is adjusted. That is, the prices will be flipped to conform to the queue shown in Figure 1.
        
        \subsection{Calculating the Price}
            Using the definition of relative price as a building block for absolute price, we have established a system to generate alternatives in calculating the price of a pair. Assume we want the price of the two assets A / E, then as per the definition visualized in Figure 1, the potential source nodes have the form $s = A / a$, and the targets have the form $t = b / E$ where assets $a$ and $b$ are arbitrary. This insight provides the foundation of randomizing source and target pools.

            A modified version of Dijkstra's algorithm \cite{dijkstra2022note} will be used to find the shortest path from the source pool to the target pool. Because the graph is composed of edge weights \( E_{n,t} = I_{n,t} \) the shortest path would yield the queue of pools that are most correlated with each other. The product of pool prices on the shortest path between the randomly chosen source and target, where the path is a compatible queue as explained in Figure 1, will yield the price of starting node's numerator asset to ending node's denominator asset, $A/E$.
        
        Because this way of pathfinding is distance minimizing on a graph where edge weights represent irregularity, the path chosen will always favor correlated pairs. The modification of the algorithm primarily stems from the fact that we need to ensure the numerator of the posterior should equal the denominator of the prior to ensure intermediary assets cancel out when multiplied, as shown in Fig 1. Djisktra's modified version, \lstinline{selective_dijkstra()} checks if the probable posterior node is compatible to the current node by comparing their numerator and denominator respectively. 
            
\begin{lstlisting}
def selective_dijkstra(self, src):

    dist = [1e7] * self.V
    dist[src] = 0
    sptSet = [False] * self.V

    for cout in range(self.V):
        u = self.minDistance(dist, sptSet)
        sptSet[u] = True

        for v in range(self.V):
            if (self.graph[v.denominator] == self.graph[v.numerator] and
            self.graph[u][v] > 0 and
            sptSet[v] == False and
            dist[v] > dist[u] + self.graph[u][v]):
                dist[v] = dist[u] + self.graph[u][v]
\end{lstlisting}

    \section{Discussion}

        \subsection{Significance}
            In this paper, we hypothesized the algorithm with the outlined design and implementation details can produce price information as accurate as Centralized Exchange prices. The proposed algorithm can provide accurate price feeds of a security by taking potentially dishonest price information of the market as an input. This advancement in the algorithmic pricing of currencies provides an alternative way to query the price of a currency without needing to trust a third party. The trustless nature of this approach is especially important in complimenting other trustless systems, such as decentralized applications hosted on public blockhains.

        \subsection{Limitations}
            \textbf{Edge Manipulation}: Despite the manipulation evasive approach used in the algorithm, if the starting pool or target pool are manipulated, the path of prices will get manipulated regardless. We assume the attacker is a rational actor, so the attacker will choose the set of source pools or target based on the cumulative liquidity of pool options. The attacker will compare the sum of liquidity of the set of source pools and the set of target pools, and will manipulate the set of pools that have the lowest sum of liquidity.

    \section{Conclusion}

    \bibliographystyle{unsrt}
    \bibliography{reference}

\end{document}